\documentclass[preprint,pre,nobibnotes,twocolumn,10pt]{revtex4}%
\usepackage{amsfonts}
\usepackage{amsmath}
\usepackage{amssymb}
\usepackage{graphicx}%
\providecommand{\U}[1]{\protect\rule{.1in}{.1in}}

\begin{document}
\preprint{ }
\preprint{UATP/1806}
\title{Hybrid Einstein-Langevin Approach for Microscopic formulation of Viscous
Drag:\ An Alternative to the Langevin Equation}
\author{P.D. Gujrati,$^{1,2}$ }
\affiliation{$^{1}$Department of Physics, $^{2}$Department of Polymer Science, The
University of Akron, Akron, OH 44325}
\email{pdg@uakron.edu}

\begin{abstract}
We present a \emph{novel hybrid but thermodynamic approach} to provide an
alternative to the Langevin equation by using system-intrinsic (SI) microwork
$d_{\text{i}}W_{k,\text{BP}}$ done by the Brownian particle (BP) in the $k$th
microstate (realization) $\mathfrak{m}_{k}$. The corresponding SI-microforce
$\mathbf{F}_{k,\text{BP}}$ is unique to $\mathfrak{m}_{k}$ and determines the
microscopic equation of motion for it. Being a thermodynamic approach, the
equipartition theorem is always satisfied and no additional stochastic
Langevin force is needed. We determine instantaneous and long-time averages of
useful quantities and thus provide a new unified approach to the fluctuating
motion from mesoscopic to macroscopic scales.

\end{abstract}
\date{\today}
\maketitle

The behavior of a microscopically-visible Brownian particle (about a micron in
diameter such as human red blood cell), first successfully formulated by
Einstein \cite{Einstein-BrownianMotion}, is perhaps the easiest prototypical
behavior that appears in many nonequilibrium (NEQ) systems, where we encounter
nonuniformity due to length scales of its constituents. Einstein assumed that
a Brownian particle (BP) can be simply described by its stochastic
\emph{center of mass }position\emph{ }$\mathbf{r}$\emph{\ }for its
specification and by ignoring the center of mass momentum and the
specification of its constituent atoms or molecules. The diffusion of the BP
obeys a diffusion equation, i.e., a Fokker--Planck equation describing
stochasticity in the ensemble picture \cite{Keizer}, which Einstein solved in
equilibrium (EQ). Langevin \cite{Langevin} later provided a very different
formulation of the same motion as a stochastic process by applying Newton's
second law $Md^{2}\mathbf{r}/dt^{2}=\mathbf{F}(t)$ to the BP of mass $M$, by
dividing $\mathbf{F}$ into a \emph{systematic} viscous force $\mathbf{F}%
_{\text{f}}(t)$ and a stochastic force $\boldsymbol{\xi}(t)$ to each
particular realization of the Brownian motion; see Chandrasekhar
\cite{Chandrasekhar} for an elegant discussion and inherent assumptions.

The distinct approaches\ by Einstein and Langevin have developed into
mathematically distinct but physically equivalent ways (Fokker-Planck versus
Langevin) to investigate stochastic processes \cite{Keizer}. The approach by
Einstein adopts a probabilistic approach to capture thermodynamic
stochasticity and results in ensemble averages such as the root-mean-square
displacement but dynamics is not a central issue. In contrast, Langevin's
approach starts with the dynamical equation in which $\mathbf{F}_{\text{f}%
}(t)$ is a thermodynamic average systematic force associated with dissipation
(thus, satisfying the second law), while stochasticity is controlled by
$\boldsymbol{\xi}(t)$ having a probability distribution of a $\delta
$-correlated \emph{white Gaussian} noise determined by its first two moments
\cite{Chandrasekhar}; they are its zero mean and constant standard deviation.
This means that the Langevin force $\boldsymbol{\xi}(t)$ defines a stationary
process because the probability distribution does not change in time, a
well-known property of white noise. In the absence of the stochastic force,
the velocity vanishes as $t\rightarrow\infty$ due to the viscous force.
Langevin used this observation to justify the inclusion of the stochastic
force \cite{Langevin}. Such a separation is one of the basic assumptions as
discussed by Chandrasekhar \cite{Chandrasekhar}, \cite{Mazur}, and Pomeau and
Piasecki \cite{Pomeau}.

It is not surprising that the Langevin equation now plays a dominant role in
the development of the modern microscopic nonequilibrium (NEQ) stochastic
thermodynamics \cite{Sekimoto,Sekimoto-Book,Seifert,Seifert-Rev}, where
attempt has been to \emph{extend} the (macroscopic) first law of
thermodynamics to the level of \emph{microstates}. (All quantities at the
microstate level are called \emph{microquantities} as opposed to their
ensemble averages, which we call \emph{macroquantities}.) This requires
introducing the concept of microwork and microheat for a microstate
\cite{Gujrati-GeneralizedWork,Gujrati-GeneralizedWork-Expanded}; their
ensemble averages will be called macrowork or simply work and macroheat or
simply heat in the macroscopic NEQ thermodynamics (NEQT)
\cite{deGroot,Prigogine,Coleman}. The concept of microwork and microheat seems
to play a central role at diverse length scales from mesoscopic to macroscopic
lengths
\cite{Bochkov,Jarzynski,Crooks,Pitaevskii,Sekimoto,Sekimoto-Book,Seifert,Lebowitz,Alicki}%
; see also some recent reviews \cite{Seifert-Rev,Maruyama}.

The Langevin equation in one dimension is given by%
\begin{equation}
Mdv(t)/dt=F(t)=-\gamma v(t)+\xi(t), \label{LangevinEquation}%
\end{equation}
with $\gamma>0$ in the systematic force to give rise to \emph{dissipation}
according to the second law, and $\xi(t)$ is the Gaussian white force
independent of the state of the system \cite{Mazur}. Both $\gamma$ and
$\xi(t)$ are \emph{independent} of the position and velocity of the BP so the
two forces are independent despite arising from the interaction of the BP with
its surroundings. Chandrasekhar \cite{Chandrasekhar} emphasizes $\xi(t)$ as a
\emph{characteristic} of a BP, which undergoes rapid fluctuations over an
interval $\Delta t$ over which $v$ only undergoes a small variation. Assuming
Stokes' law for a spherical Brownian particle of radius $a$, we have
$\gamma=6\pi a\eta>0$, where $\eta$ is the viscosity of the surrounding fluid.
It follows that we can identify a "linear dimension" $l$ of a BP so that we
can write $\gamma=\eta l$. In the stochastic energetics proposed by Sekimoto
\cite{Sekimoto,Sekimoto-Book}, this equation is taken to apply to each
realization (microstate) of the Brownian motion, with $\gamma$ still
satisfying $\gamma>0$.

In the absence of $\xi(t)$, and with $v_{0}$ as the initial velocity $v(0)$,
we have%
\begin{equation}
v(t)=v_{0}e^{-\gamma t/M}\rightarrow0\text{ as }t\rightarrow\infty\text{
}(\gamma>0), \label{Velocity-Simple}%
\end{equation}
whereas in equilibrium ($t\rightarrow\infty$) we expect $\left\langle
v^{2}\right\rangle =T/M$ from the equipartition theorem from the ensemble
average over all possible values of $v$ or for that matter $v_{0}$. Therefore
Langevin assumed the existence of $\xi(t)$, whereupon one needs to perform two
distinct and independent averages over $v_{0}$ and $\xi$ at each time.\ The
equipartition theorem is always fulfilled in Einstein's approach
\cite{Einstein-BrownianMotion} being based on equilibrium thermodynamics.

We now propose an alternative approach to study each realization
$\mathfrak{m}_{k}$ of the Brownian particle, which is a hybrid of the above
two approaches in that we do not need $\xi(t)$ but the thermodynamic
stochasticity appears due to the ensemble average as done by Einstein, and
follows the standard formulation for a statistical system \cite[for
example]{Landau} by considering the ensemble average of a quantity such as the
velocity $\left\{  \mathbf{v}_{k}\right\}  $ over various microstates
$\left\{  \mathfrak{m}_{k}\right\}  $ at each instant. Thus, $\mathbf{v}$
becomes a random variable whose outcomes are $\{\mathbf{v}_{k}\}$ on
$\{\mathfrak{m}_{k}\}$. As microstate probabilities $p_{k}$ continue to
change, the ensemble average is not stationary as for the white Langevin force.

The microforce $F_{k}$, see Eq. (\ref{MicroForce-Velocity}), associated with
$\mathfrak{m}_{k}$ is the outcome of some random microforce and the
macroforce, the ensemble average, is the systematic force of a particular sign
in the Langevin equation. The non-vanishing fluctuations, see Eqs.
(\ref{F-Fluctuation0},\ref{F-Fluctuation}), in $F_{k}$ even in equilibrium
demonstrates that these outcomes have \emph{both} signs. However, instead of
considering $F_{k}$, we consider the microwork $d_{\text{i}}W_{k}$\ done by
it, whose average $d_{\text{i}}W\geq0$ ($\gamma>0$)\ in accordance with the
second law. The microwork $d_{\text{i}}W_{k}$, being specific to
$\mathfrak{m}_{k}$, has a unique value that is independent of $p_{k}$ but
changes over microstates.

Here we take a major departure from Langevin's approach by not restricting the
\emph{sign} of $d_{\text{i}}W_{k}$ over all microstates for the simple reason
that the second law is not applicable at the level of microstates; the law
only emerges as the ensemble average is taken as is well known; see also
\cite{Gujrati-GeneralizedWork,Gujrati-GeneralizedWork-Expanded} for a clear
demonstration. In the context of the Langevin equation, this means that
$\gamma$ will be of either sign so that for some microstate, $v(t)$ in Eq.
(\ref{Velocity-Simple}) may decrease and go to zero, while for others, it may
increase in magnitude and diverge to infinity, as $t\rightarrow\infty$, with
the condition that $\left\langle v^{2}\right\rangle $ will satisfy the
equipartition theorem as we will demonstrate. We use the version NEQT that we
have developed recently
\cite{Gujrati-GeneralizedWork,Gujrati-GeneralizedWork-Expanded} and briefly
discussed below. We also derive the equations of motion using Newton's second
law; see Eqs. (\ref{NewLangevinEq},\ref{NewLangevinEqAv}).
\begin{figure}
[ptb]
\begin{center}
\includegraphics[
height=1.9657in,
width=3.8207in
]%
{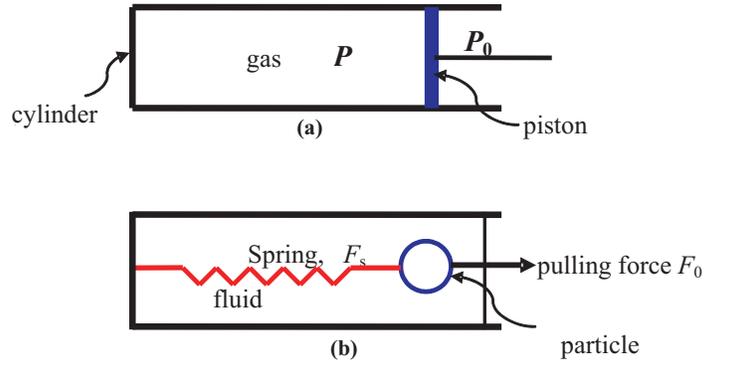}%
\caption{We schematically show a system of (a) gas in a cylinder with a
movable piston under an external pressure $P_{0\text{ }}$controlling the
volume $V$ of the gas, and (b) a particle attached to a spring in a fluid
being pulled by an external force $F_{0}$, which causes the spring to stretch
or compress depending on its direction. In an irreversible process, the
internal pressure $P$ (the spring force $F_{\text{s}}$) is different in
magnitude from the external pressure $P_{0}$ (external force $F_{0}$).}%
\label{Fig_Piston-Spring}%
\end{center}
\end{figure}

We consider the two examples (a) and (b) in Fig. \ref{Fig_Piston-Spring} as
our system $\Sigma$\ and treat the piston or the particle as a Brownian
particle (BP) of mass $M$. We assume that the piston in (a) may be either
mesoscopic or microscopic, while the particle in (b) will be assumed to denote
a mesoscopic particle of micron-like size. We will establish in our approach
that both experience fluctuating Brownian motion, except that for the
macroscopic size piston, it is not noticeable because of its macroscopic mass.
We follow Einstein and focus on the center-of-mass (CM)\ of the BP, and denote
rest of $\Sigma$ by excluding the BP by $\Sigma_{\text{R}}$. Let $V$ denote
the volume of $\Sigma$ and $\mathbf{P}_{\text{BP}}$ and $\mathbf{P}_{\text{R}%
}$\textbf{\ }the linear momenta of the BP and $\Sigma_{\text{R}}$,
respectively. We assume that $\Sigma$ is stationary in the lab-frame so that%
\begin{equation}
\mathbf{P}_{\text{BP}}+\mathbf{P}_{\text{R}}=0;
\label{Stationary_Momentum_Condition}%
\end{equation}
the medium $\widetilde{\Sigma}$ is also stationary. We will treat the piston
problem for simplicity as it is commonly discussed in introductory physics.
Let $\mathbf{x}$ denote a phase point in the phase space of $\Sigma$ so that
the Hamiltonian of the system is written as $\mathcal{H}(\left.
\mathbf{x}\right\vert V,\mathbf{P}_{\text{BP}},\mathbf{P}_{\text{R}})$ in
which $V,\mathbf{P}_{\text{BP}}$ and $\mathbf{P}_{\text{R}}$ appear as
parameters: the variations $dV,d\mathbf{P}_{\text{BP}}$ and $d\mathbf{P}%
_{\text{R}}$ change the value of $\mathcal{H}$; this change represents the
\emph{generalized} work $dW$ done by the system as shown elsewhere
\cite{Gujrati-Heat-Work0,Gujrati-Heat-Work,Gujrati-I,Gujrati-II,Gujrati-III,Gujrati-Entropy1,Gujrati-Entropy2}%
. Introducing the system-intrinsic (SI) "mechanical forces" obtained directly
from the SI-Hamiltonian%
\begin{subequations}
\begin{equation}
P_{\mathbf{x}}\doteq-\frac{\partial\mathcal{H}(\mathbf{x})}{\partial
V},-\mathbf{V}_{\mathbf{x,}\text{BP}}\doteq-\frac{\partial\mathcal{H}%
(\mathbf{x})}{\partial\mathbf{P}_{\text{BP}}},-\mathbf{V}_{\mathbf{x,}%
\text{R}}\doteq-\frac{\partial\mathcal{H}(\mathbf{x})}{\partial\mathbf{P}%
_{\text{R}}} \label{Microforces}%
\end{equation}
for each realization denoted by $\mathbf{x}$, which we call microforces here,
we can identify the corresponding microworks by $P_{\mathbf{x}}dV$, etc. so
that the net SI-microwork becomes
\begin{equation}
dW_{\mathbf{x}}=P_{\mathbf{x}}dV-\mathbf{V}_{\mathbf{x,}\text{BP}%
}\mathbf{\cdot}d\mathbf{P}_{\text{BP}}-\mathbf{V}_{\mathbf{x,}\text{R}%
}\mathbf{\cdot}d\mathbf{P}_{\text{R}} \label{Microwork}%
\end{equation}
The ensemble averages of the various microwork components are given by $PdV$,
etc., see Landau and Lifshitz \cite{Landau,Landau-Fluid} and elsewhere
\cite[and references theirin]{Gujrati-II}, where%
\begin{equation}
P\doteq-\partial E/\partial V,\mathbf{V}_{\text{BP}}\doteq\partial
E/\partial\mathbf{P}_{\text{BP}},\mathbf{V}_{\text{R}}\doteq\partial
E/\partial\mathbf{P}_{\text{R}} \label{Macroforces}%
\end{equation}
denote average "SI-forces"; here $E$ denotes the ensemble average energy in
the lab frame: $E=\left\langle \mathcal{H}(\left.  \mathbf{x}\right\vert
V,\mathbf{P}_{\text{BP}},\mathbf{P}_{\text{R}})\right\rangle $ over all
realizations $\mathbf{x}$, and $P$ is the average pressure and $\mathbf{V}%
_{\text{BP}},\mathbf{V}_{\text{R}}$ are the average velocities of the BP and
of $\Sigma_{\text{R}}$, respectively. It is clear that $E$ depends on the
parameters $V,\mathbf{P}_{\text{BP}},\mathbf{P}_{\text{R}}$ that determine
various microforces in Eq. (\ref{Macroforces}). Accordingly, the system
entropy, which we assume is a \emph{state} function, is written as
$S(E,V,\mathbf{P}_{\text{BP}},\mathbf{P}_{\text{R}})$ so that the system is in
internal equilibrium (IEQ) \cite{Gujrati-I,Gujrati-II} but not in equilibrium
with the medium; the temperature and pressure of $\widetilde{\Sigma}$\ are
denoted by $T_{0}$ and $P_{0}$, respectively. The temperature $T$ of the
system is defined as $T\doteq\partial E/\partial S$. The state entropy
$S(E,V,\mathbf{P}_{\text{BP}},\mathbf{P}_{\text{R}})$ in internal equilibrium
has the maximum possible value for given $E,V,\mathbf{P}_{\text{BP}}$, and
$\mathbf{P}_{\text{R}}$. It then follows, as we have shown earlier
\cite{Gujrati-I,Gujrati-II,Gujrati-III,Gujrati-Entropy1,Gujrati-Entropy2} that
most of the results from equilibrium statistical mechanics will also apply to
the system in IEQ at its temperature and pressure $T$ and $P$, respectively,
in a larger state space; see later also, which explains the importance of
internal equilibrium. The generalized work $dW$ in terms of average forces is
\begin{equation}
dW=PdV-\mathbf{V}_{\text{BP}}\mathbf{\cdot}d\mathbf{P}_{\text{BP}}%
-\mathbf{V}_{\text{R}}\mathbf{\cdot}d\mathbf{P}_{\text{R}} \label{dW-relation}%
\end{equation}
The Gibbs fundamental relation for $E$ is given by $dE=TdS-PdV+\mathbf{V}%
_{\text{BP}}\mathbf{\cdot}d\mathbf{P}_{\text{BP}}+\mathbf{V}_{\text{R}%
}\mathbf{\cdot}d\mathbf{P}_{\text{R}}$. Using Eq.
(\ref{Stationary_Momentum_Condition}), we can rewrite this equation as
$dE=TdS-PdV+\mathbf{V\cdot}d\mathbf{P}_{\text{BP}}$ in terms of the
$\emph{relative}$ \emph{velocity}, also known as the \emph{drift velocity
}$\mathbf{V\doteq V}_{\text{BP}}-\mathbf{V}_{\text{R}}$ of the BP with respect
to $\Sigma_{\text{R}}$. We can also rewrite the drift velocity term using
$\mathbf{V\cdot}d\mathbf{P}_{\text{BP}}\equiv\mathbf{F}_{\text{BP}%
}\mathbf{\cdot}d\mathbf{R}$, where $\mathbf{F}_{\text{BP}}\doteq
d\mathbf{P}_{\text{BP}}\mathbf{/}dt$ is the average \emph{force}\ and
$d\mathbf{R=V}dt$ is the average \emph{relative displacement} of the BP to
provide a justification for Einstein's approach using the CM of the BP and not
considering its momentum at all.

The generalized macrowork is now $dW=PdV-\mathbf{F}_{\text{BP}}\mathbf{\cdot
}d\mathbf{R}$ and the generalized macroheat is $dQ=TdS$. We can now write
$dE=dQ-dW$, which expresses the first law in terms of the generalized
quantities. This expresses an important fact: the two terms in it denote\emph{
independent} variations of the energy $E$: $dQ$ denotes the change due to
entropy variation and $dW$ isentropic variation (constant entropy). This
allows us to deal with $dW$ as a purely mechanical ($dS=0$) quantity resulting
in microstate energies, and $dQ$ as due to variations in microstate
probabilities \cite{Gujrati-GeneralizedWork,Gujrati-GeneralizedWork-Expanded}.
which we summarize below for the benefit of the reader. For convenience, we
treat $\mathbf{x}$\ as discrete and use $k$ for it in the following. The
average energy $E$ is defined in terms of microstate energies $E_{k}%
=E_{k}(V,\mathbf{P}_{\text{BP}})$ [or equivalently $E_{k}=E_{k}(V,\mathbf{R}%
)$] and microstate probabilities $p_{k}$ as an ensemble average $\left\langle
E\right\rangle $, written simply as $E\doteq%
{\textstyle\sum\nolimits_{k}}
E_{k}p_{k},$ so that $dE=%
{\textstyle\sum\nolimits_{k}}
E_{k}dp_{k}+%
{\textstyle\sum\nolimits_{k}}
p_{k}dE$, where $dE_{k}=(\partial E_{k}/\partial V)dV+(\partial E_{k}%
/\partial\mathbf{P}_{\text{BP}})\mathbf{\cdot}d\mathbf{P}_{\text{BP}}$;
compare with Eqs. (\ref{Microforces}) and (\ref{Microwork}) after replacing
$\mathbf{x}$ with $k$. (Recall that we can also use $\mathbf{R}$ instead of
$\mathbf{P}_{\text{BP}}$\ as an argument in $E_{k}$.) The first sum in $dE$
involves $dp_{k}$ at fixed $E_{k}$, and evidently corresponds to the entropy
change $dS$. This sum denotes the generalized heat $dQ$. The second sum
involves $dE_{k}$ at fixed $p_{k}$, and evidently corresponds to $dS=0$. As
$dE_{k}$ is due to parameter changes in the Hamiltonian, it is related to the
generalized work $(-dW_{k})$, which is the analog of Eq. (\ref{dW-relation})
for microstates. Its average gives rise to $(-dW)$, which then justifies the
above rendition of the first law.

The exchange work is $d_{\text{e}}W=P_{0}dV$ and the irreversible work is
$d_{\text{i}}W\doteq dW-d_{\text{e}}W$ associated with $\Sigma$ is%
\end{subequations}
\begin{equation}
d_{\text{i}}W=(P-P_{0})dV-\mathbf{F}_{\text{BP}}\mathbf{\cdot}d\mathbf{R\geq
}0. \label{Irreversible Work}%
\end{equation}
The inequality is in accordance with the second law
\cite{Gujrati-GeneralizedWork,Gujrati-GeneralizedWork-Expanded}. Similarly,
the exchange heat with $\widetilde{\Sigma}$\ is $d_{\text{e}}Q=T_{0}%
d_{\text{e}}S$ and the irreversible heat is $d_{\text{i}}Q\doteq
dQ-d_{\text{e}}Q=(T-T_{0})d_{\text{e}}S+Td_{\text{i}}S~\mathbf{\geq}0$. Again,
the last inequality is in accordance with the second law
\cite{Gujrati-GeneralizedWork,Gujrati-GeneralizedWork-Expanded}. As the first
law can also be written as $dE=d_{\text{e}}Q-d_{\text{e}}W$, we must have
$d_{\text{i}}Q=d_{\text{i}}W$ in magnitude, even though one is caused by
changes in the probabilities and the other one by changes in microstate energies.

The irreversible contributions in $d_{\text{i}}W$ are due to some kind of
"force" imbalance as pointed out recently
\cite{Gujrati-GeneralizedWork,Gujrati-GeneralizedWork-Expanded}. Away from
equilibrium, $P\neq P_{0}$ so the pressure imbalance $\Delta P\doteq P-P_{0}$
within $\Sigma$ determines the the first irreversible contribution in
$d_{\text{i}}W$ above. Similarly, the average force and the relative velocity
vanish in equilibrium ($\mathbf{F}_{\text{BP}}=0,\mathbf{V}=0$) so away from
equilibrium, $\mathbf{F}_{\text{BP}}$ or $\mathbf{V}$ represents the force
imbalance or the relative velocity imbalance and determines the second
irreversible contribution $-\mathbf{F}_{\text{BP}}\mathbf{\cdot}d\mathbf{R}$
or $-\mathbf{V\cdot}d\mathbf{P}_{\text{BP}}$ in $d_{\text{i}}W$. From now on,
we will assume $F_{0}$ and $F_{\text{s}}$ to be identically zero for
simplicity for the particle as we wish to pursue the consequences of the
relative velocity.

It follows from the second law that each term on the right side in Eq.
(\ref{Irreversible Work}) must be nonnegative so that $d_{\text{i}%
}W_{\text{f,BP}}\equiv-\mathbf{V\cdot}d\mathbf{P}_{\text{BP}}=-\mathbf{F}%
_{\text{BP}}\mathbf{\cdot}d\mathbf{R}\geq0$; here, f is for friction. Hence,
we can express $\mathbf{F}_{\text{BP}}=-\mathbf{V}f(V^{2},t)$, where
$f(V^{2},t)>0$ is an even function of $\mathbf{V}$\ at each instant so that
$d_{\text{i}}W_{\text{f,BP}}=f(V^{2},t)\mathbf{V\cdot}d\mathbf{R}\geq0$. To
make connection with the Langevin equation, we will assume $f$ to be a power
series with $f(0,t)=\gamma(t)$, which Langevin takes to be a constant so that
$d_{\text{i}}W_{\text{f,BP}}\simeq\gamma\mathbf{V\cdot}d\mathbf{R}$ is the
\emph{frictional work} in a small-velocity approximation. The above discussion
provides a thermodynamic justification of the viscous drag in the Langevin equation.

It follows from Eqs. (\ref{Microforces}) and (\ref{Macroforces}) that the
microanalog of $P$ is $P_{k}$, which differs from $P$ and the fluctuations
$\Delta P\doteq P_{k}-P$ determine the mean square fluctuation $\left\langle
(\Delta P)^{2}\right\rangle >0$
\cite{Gujrati-GeneralizedWork,Gujrati-GeneralizedWork-Expanded}. It is known
from equilibrium statistical mechanics that this fluctuation $\left\langle
(\Delta P)^{2}\right\rangle _{\text{eq}}=-T(\partial P/\partial V)_{S}$ in EQ
for $P=P_{0}$ is not identically zero \cite{Landau} so $P_{k}\neq P$ in
general. Similarly, there are fluctuations in $\mathbf{F}_{k\text{,BP}}$ or
$\mathbf{V}_{k\text{,BP}}$\ (the analog of $P_{k}$) around the average
$\mathbf{F}_{\text{BP}}$ or $\mathbf{V}_{\text{BP}}$, which are always
present; see Eqs. (\ref{F-Fluctuation0}-\ref{F-Fluctuation}). In equilibrium,
the average force and the average relative velocity vanish: $\mathbf{F}%
_{\text{BP,eq}}=0,\mathbf{V}_{\text{eq}}=0$, but there are fluctuations in
their microvalues from microstate to microstate even in equilibrium as noted
above. These fluctuations are the hallmark of a statistical system and must be
accounted for whether we consider a reversible or an irreversible process.

The significance of the irreversible work $(P-P_{0})dV$ is well known and has
also been discussed elsewhere \cite[and references theirin]{Gujrati-II}. Here,
we will consider a free BP ($F_{0}=0,F_{\text{s}}=0$) for which we are
interested in studying the dissipation $d_{\text{i}}W_{f,\text{BP}}$ due to
friction generated by the relative motion $\mathbf{V}$; the friction finally
brings about the EQ macrostate with $\mathbf{F}_{\text{BP,eq}}=0$ or
$\mathbf{V}_{\text{eq}}=0$. As said above, there are still force or velocity
fluctuations both for the piston and the Brownian particle, having different
length scales. Thus, our approach unifies the two different length scales.

We will suppress the suffix f on $d_{\text{i}}W_{\text{f,BP}}$ for simplicity
now. The irreversible work $d_{\text{i}}W_{\text{BP}}\simeq\gamma
(t)\mathbf{V}(t)\mathbf{\cdot}d\mathbf{R}(t)=\gamma(t)V^{2}(t)dt\geq0$ at a
given instant $t$ is an average over all microstates at that instant. We can
infer from it the form of the internal microwork (suffix $k$) as $d_{\text{i}%
}W_{k\text{,BP}}=-\mathbf{F}_{k,\text{BP}}(t)\mathbf{\cdot}d\mathbf{R}%
(t)=-\mathbf{V}_{k}(t)\mathbf{\cdot}d\mathbf{P}_{\text{BP}}(t)$ associated
with the microstate $k$. In terms of $E_{k}$, we have
\begin{equation}
\mathbf{F}_{k\text{,BP}}\doteq-\partial E_{k}/\partial\mathbf{R}\text{ or
}\mathbf{V}_{k\text{,BP}}\doteq-\partial E_{k}/\partial\mathbf{P}_{\text{BP}}.
\label{MicroForce-Velocity}%
\end{equation}
The important point is that this internal work has no sign restriction. This
is our main point of departure from Langevin. Our equation of motion for the
BP in $\mathfrak{m}_{k}$ is
\begin{subequations}
\begin{equation}
md^{2}\mathbf{R}_{k}(t)/dt^{2}=\mathbf{F}_{k,\text{BP}}(t),
\label{NewLangevinEq}%
\end{equation}
where $m$ is the reduced mass of the BP. The stochasticity emerges as we
average this equation over all microstates using $p_{k}$; the result is%
\begin{equation}
md^{2}\mathbf{R}(t)/dt^{2}=\mathbf{F}_{\text{BP}}(t), \label{NewLangevinEqAv}%
\end{equation}
with $\mathbf{F}_{\text{BP}}(t)$ playing the role of the systematic (or
average) force. The difference $\Delta\mathbf{F}_{k,\text{BP}}(t)\doteq
\mathbf{F}_{k,\text{BP}}(t)-\mathbf{F}_{\text{BP}}(t)$ seems to resemble
$\boldsymbol{\xi}(t)$. This is where another important difference from the
Langevin approach appears in which $\boldsymbol{\xi}(t)$ takes all possible
values for each realization. In our approach, there is only one unique value
of $\mathbf{F}_{k\text{,BP}}\doteq-\partial E_{k}/\partial\mathbf{R}$ for
$\mathfrak{m}_{k}$ so $\Delta\mathbf{F}_{k,\text{BP}}(t)$ also takes a
\emph{single} value on it. It changes its value over different $\mathfrak{m}%
_{k}$'s.

We can use the standard fluctuation theory
\cite{Gujrati-Entropy2,Landau,Gujrati-Fluctuations} to obtain
\emph{instantaneous} fluctuations in $\mathbf{F}_{\text{BP}},\mathbf{R,V}$ and
$\mathbf{P}_{\text{BP}}$ when the system is in the IEQ state state. We
restrict ourselves to a $1$-d case for simplicity ($R$ replaced by $x$). The
conclusion is that the probability of fluctuations about the IEQ state is
given by $W_{0}\exp(-\beta\rho/2)$, where $\rho=\Delta T\Delta S-\Delta
P\Delta V+\Delta F_{\text{BP}}\Delta x$ in terms of various fluctuations. We
will use the approximation that $\partial P/\partial F_{\text{BP}}$ vanishes.
This ensures that the fluctuations in $T,V$ and $F_{\text{BP}}$ are
independent. The results for square fluctuations involving $T$ and $V~$are
already known \cite{Landau} so here we only focus on the remainder
fluctuations due to $F_{\text{BP}}$. We easily find that the coefficient of
$(\Delta F_{\text{BP}})^{2}$ in $\rho$\ is $(\partial x/\partial F_{\text{BP}%
})_{T,V}$, the derivative taken in the IEQ state. It then follows from the
fluctuation theory that%
\end{subequations}
\begin{subequations}
\begin{equation}
\langle(\Delta F_{\text{BP}})^{2}\rangle=T(\partial F_{\text{BP}}/\partial
x)_{T,V}. \label{F-Fluctuation0}%
\end{equation}
Observing from Eq. (\ref{MicroForce-Velocity}) that $F_{\text{BP}}(V,x)$ is a
function of $x$, and using $F_{\text{BP}}\simeq-\gamma\overset{\cdot}{x}$, we
find that $\partial F_{\text{BP}}/\partial x=-\gamma\overset{\cdot\cdot}%
{x}/\overset{\cdot}{x}=\gamma^{2}/m$ so that
\begin{equation}
\left\langle (\Delta F_{\text{BP}})^{2}\right\rangle =T\gamma^{2}/m=Tl^{2}%
\eta^{2}/m>0, \label{F-Fluctuation}%
\end{equation}
which is precisely what we expect in this approximation since $(\Delta
F_{\text{BP}})^{2}=\gamma^{2}\overset{\cdot}{x}^{2}$and $\langle\overset
{\cdot}{x}^{2}\rangle$ $=T/m$ as shown below. In equilibrium, $\Delta
F_{k,\text{BP}}=F_{k,\text{BP}}$ so $F_{k,\text{BP}}$ takes all possible
values of both signs. As the values of $\{F_{k,\text{BP}}\}$ are intrinsic to
$\{\mathfrak{m}_{k}\}$, these values remain the same in any macrostate. We can
similarly obtain $\left\langle (\Delta x)^{2}\right\rangle =T(\partial
x/\partial F_{\text{BP}})_{T,V}=mT/\gamma^{2}=mT/l^{2}\eta^{2}>0$. In a highly
viscous environment, the mean square CM-fluctuation becomes very small as
expected, and $\langle(\Delta F_{\text{BP}})^{2}\rangle$ become large. All
these results are valid for any BP of any size (linear dimension $l$) ranging
from mesoscales to macroscales. We can use the standard fluctuation theory
\cite{Landau,Gujrati-Fluctuations} to obtain instantaneous fluctuations in
$\mathbf{F}_{\text{BP}},\mathbf{R,V}$ and $\mathbf{P}_{\text{BP}}$, which is a
standard calculation but we will not stop here to that.

We turn to the important aspects of our approach. As shown elsewhere
\cite{Gujrati-GeneralizedWork,Gujrati-GeneralizedWork-Expanded} and also
mentioned above by the definition of microforces and microworks in Eq.
(\ref{Microforces}), $dW_{k}=-dE_{k}$ in general. For the free BP, this
reduces to $d_{\text{i}}W_{k}=-d_{\text{i}}E_{k}$, where $d_{\text{i}}E_{k}$
is the change in the microstate energy due to internal processes due to force
imbalance, \textit{i.e.} due to $\Delta P_{k}$ and $\mathbf{F}_{k\text{,BP}}$.
Here, we will not be concerned with $\Delta P_{k}$. Hence, we can use
$d_{\text{i}}W_{k\text{,BP}}$ to determine the change $\Delta_{\text{i}%
}E_{k\text{,BP}}$ for the BP over an interval $(0,t)$. We have%
\end{subequations}
\[
\Delta_{\text{i}}E_{k\text{,BP}}=%
{\textstyle\int\nolimits_{0}^{t}}
\mathbf{F}_{k,\text{BP}}(t)\mathbf{\cdot}d\mathbf{V}_{k}(t)dt,
\]
where we have set $\mathbf{V}_{k}(t)\doteq d\mathbf{R}(t)/dt$ as the velocity
for $\mathfrak{m}_{k}$. Using $\mathbf{F}_{k,\text{BP}}(t)=md\mathbf{V}%
_{k}(t)/dt$, we have
\[
\Delta_{\text{i}}E_{k\text{,BP}}=(m/2)(\mathbf{V}_{k}^{2}(t)-\mathbf{V}%
_{k}^{2}(0)),
\]
which is nothing but the change in the kinetic energy of the center of mass of
the BP. This is nothing but the work-energy theorem from classical mechanics.

The equation of motion for a given microstate now becomes in this
approximation%
\begin{equation}
d\mathbf{V}_{\text{$k$}}(t)/dt=-(\gamma_{k}(t)/m)\mathbf{V}_{k}(t),
\label{NewLangevinEq-micro}%
\end{equation}
whose solution is%
\[
\mathbf{V}_{\text{$k$}}(t)=\mathbf{V}_{\text{$k$}}(0)\exp(-%
{\textstyle\int\nolimits_{0}^{t}}
\gamma_{\text{$k$}}(u)du/m).
\]
We see that the components of the possible velocities range from $-\infty$ to
$+\infty$. We van now evaluate the average of $V_{k}^{2}(t)$ at each instant,
assuming as we have done that the system is in internal equilibrium. This
means that the velocity distribution is given by the Maxwell distribution at
temperature $T$ \cite{Einstein-BrownianMotion,Chandrasekhar} so we have the
standard result%
\[
\left\langle \mathbf{V}^{2}(t)\right\rangle =3T(t)/m,
\]
where $T(t)$ is the instantaneous temperature of $\Sigma$. This result can
also be directly deduced from $\left\langle (\Delta F_{\text{BP}}%
)^{2}\right\rangle \simeq\gamma^{2}\langle\overset{\cdot}{x}^{2}\rangle$. The
difference of the above conclusion with that by Langevin lies in the fact that
in our approach, $\gamma_{\text{k}}(t)$ for a microstate depends on the
microstate and has no sign restriction. Because of this, it cannot be taken
out of the averaging process. We thus see that our approach has allowed the
equipartition theorem to remain valid at all times provided $\Sigma$ is in
internal equilibrium. The stochasticity of the Brownian motion has been
captured in the approach. We see that $\left\langle \mathbf{V}^{2}%
(t)\right\rangle \propto1/m$ so larger the mass, smaller the mean square
fluctuations over time such as for a macroscopic piston. However, for a
mesoscopic Brownian particle, it can be appreciable and can be observed.

We now determine the average square displacement of the BP. For this, we
follow Einstein \cite{Einstein-BrownianMotion}\ again and recall that the
distribution function of the relative displacement $\mathbf{R}$ is given by
$f(\mathbf{R},t)=e^{-\mathbf{R}^{2}/4Dt}/(4\pi Dt)^{3/2}$, so that
\begin{equation}
\left\langle \mathbf{R}^{2}(t)\right\rangle =6Dt \label{EinsteinRelation}%
\end{equation}
as a function of time; here, $D$ is the diffusion constant, which is related
to the viscosity of the fluid by $D=T/6\pi\eta a$. We can also compute
$\left\langle \mathbf{R}^{2}(t)\right\rangle $ from\ Eq.
(\ref{NewLangevinEq-micro}) in a standard way but we will not stop \ to do that.

To summarize, we have given an alternative to the Langevin equation based on
$\mu$NEQT that was initiated a while back
\cite{Gujrati-II,Gujrati-Heat-Work,Gujrati-Entropy2}. Its usage shows that the
uniquely defined microforces and microworks in the system are, as expected,
fluctuating quantities. We make no assumptions about the nature of these
fluctuations as are needed for the stochastic forces in the Langevin equation.
We then use the microscopically deterministic equations of motion for each
realization and show that their fluctuating nature satisfies the equipartition
theorem at all times provided the system is in internal equilibrium so that
$T$ can be defined. We also reproduce the Einstein relation in Eq.
(\ref{EinsteinRelation}). The new approach differs from the Langevin approach
(LA) in

\begin{enumerate}
\item There is a \emph{unique} microforce $\mathbf{F}_{\text{f}k,\text{BP}}$
for each $\mathfrak{m}_{k}$ and requires a single averaging over $\{p_{k}\}$
to give $\mathbf{F}_{\text{f,BP}}$. In LA, there are two distinct averaging
over $\mathbf{v}_{0}$ and $\boldsymbol{\xi}$.

\item Fluctuations in $\mathbf{F}_{\text{f}k,\text{BP}}$ change with
$\{p_{k}\}$ in time and are determined by thermodynamics, while those in
$\boldsymbol{\xi}$ are stationary

\item The internal work $dW_{\text{f}k,\text{BP}}$ has no sign restriction but
the macrowork $dW_{\text{f,BP}}\geq0$. In LA, $\boldsymbol{\xi}$\ does no macrowork.

\item The approach provides a thermodynamic justification for the frictional
drag in the Langevin equation. In LA, it is taken as a fact.

\item The approach covers mesoscales to macroscales and applies to
nonequilibrium states also, while LA is limited to equilibrium states.
\end{enumerate}

\end{document}